\documentclass{article}

\usepackage[preprint,nonatbib]{neurips_2020}

\usepackage[utf8]{inputenc} 
\usepackage[T1]{fontenc}    
\usepackage{hyperref}       
\usepackage{url}            
\usepackage{booktabs}       
\usepackage{amsfonts}       
\usepackage{nicefrac}       

\usepackage{graphicx}

\usepackage{microtype}      
\usepackage{xspace}
\usepackage{bbm}
\usepackage[english]{babel}
\usepackage{amsmath,amsthm}
\usepackage{tabularx}

\usepackage{amsmath}
\usepackage{diagbox}
\usepackage{slashbox}
\usepackage{dsfont, xcolor}

\usepackage{hyperref}

\hypersetup{
    citecolor=blue,
    colorlinks=true,
    linkcolor=blue,
    filecolor=blue, urlcolor=blue,
}
\usepackage{float}

\usepackage{multirow}
\usepackage[colorinlistoftodos]{todonotes}

\usepackage{subfig}
\usepackage{algorithm}
\usepackage[noend]{algpseudocode}

\usepackage{soul}
\usepackage[style=base]{caption}
\usepackage[toc,page]{appendix}
\usepackage{mathtools}

\usepackage{amssymb}

\usepackage{hyperref}
\hypersetup{
    colorlinks,
    citecolor=black,
    filecolor=black,
    linkcolor=black,
    urlcolor=black
}

\usepackage[normalem]{ulem}

\title{AI for Investment: A Platform Disruption}

\author{%
  Mohammad Rasouli\thanks{Mohammad Rasouli is the corresponding author. All authors' biographies are put at the end of the paper.}
\\\texttt{rasoulim@stanford.edu}
\\
  \And
  Ravi Chiruvolu\\
  \texttt{ravichiruvolu@alum.mit.edu} \\
  \AND
  Ali Risheh \\
  \texttt{arisheh@calstatela.edu} \\
}

\begin{document}

\maketitle

\begin{abstract}
With the investment landscape becoming more competitive, efficiently scaling deal sourcing and improving deal insights have become a dominant strategy for funds. While funds are already spending significant efforts on these two tasks, they cannot be scaled with traditional approaches; hence, there is a surge in automating them. Many third party software providers have emerged recently to address this need with productivity solutions, but they fail due to a lack of personalization for the fund, privacy constraints, and natural limits of software use cases. Therefore, most major funds and many smaller funds have started developing their in-house \textit{AI platforms}: a game changer for the industry. These platforms grow smarter by direct interactions with the fund and can be used to provide personalized use cases. Recent developments in large language models, e.g. ChatGPT, have provided an opportunity for other funds to also develop their own AI platforms. While not having an AI platform now is not a competitive disadvantage, it will be in two years. Funds require a practical plan and corresponding risk assessments for such AI platforms.
\end{abstract}
\section{Introduction: What is driving the surge for AI in the investment business?}

With the investment landscape becoming more competitive, funds are looking to scale their deal sourcing while improving their deal insights, and LPs are looking for funds with a unique edge in this space. The economic megatrends have led to more funds and more dry powder but fewer opportunities and deals \cite{McKinsey2023_1}. Consequently, scaling the deal sourcing and improving the quality of deals is considered a dominant winning strategy \cite{Gartner2021_1, PwC2023_1}. By improving their deal sourcing, funds can get to the target asset management team faster, and by improving their deal insights, funds can ensure higher deal quality and have more focused conversations with the target management teams.

Since the traditional approach for deal sourcing and deal insight are costly and fail to scale, funds are demanding technology, particularly AI, for this purpose. Currently, the funds spend on average 75\% of their time on deal sourcing and due diligence, coming at a large opportunity cost for them to focus on value-added tasks such as relationship-building and investing \cite{PwC2023_1}. Yet, deal sourcing and deal insights cannot be scaled and improved further in traditional ways such as connecting with people, searching online, marketing campaigns, and making calls. Hence, the funds have realized the value of using automation for this purpose \cite{Gartner2021_1, Forbes2023_1}; this will also be a differentiation that can play an advantage in talking to LPs who have always looked for funds with an edge in this space: 75\% of LPs think AI is useful for deal origination, and 64\% believe the technology will have significant impact on deal assessment \cite{PEI2023_1}.

The use of alternative data for improving deal insights is another reason for funds' demand for technology solutions \cite{LS2022}. Alternative data, for example, social media, rating and review, credit spending, mobility, satellite image, key new hires, new market or product expansion, web traffic, etc., is a shift in the industry, especially in private equity and venture capital, and requires the use of artificial intelligence \cite{BlackRock2022_1}. While hedge funds and public asset investors already appreciate and incorporate the value of alternative data for generating alpha, private equities, and VCs have more recently started to experiment with them. These alternative data sources however are often massive, unstructured, and complex, which makes them impossible for a human to effectively analyze and generate insights. Fortunately, machine learning algorithms can analyze this data, pattern match, and find the signal from the noise \cite{LS2022}.

\section{Why third party software solutions fail at addressing the AI demand in investment business?}

Consequent to the above dynamics, there has been an emergence of third party solution providers in recent years to address the funds' need for using technology for investment. These third party solutions are in the form of \emph{data productivity tools} for the funds that are offered as \emph{software}. Within the space of deal sourcing and deal insight, these solutions can be categorized into three groups: a) aggregating a database of companies with their major archetypes, plus a search capability across the database. b) aggregating and processing \textit{alternative data} for deal insights and c) providing standard due diligence by organizing the process and providing ready responses to questions. 

While these third-party solutions can boost productivity, they fail to address the real needs of the funds \cite{DataDrivenVC}. This is the case for four primary reasons: Lack of personalization, limited intelligence, privacy concerns, and natural limits of software. 
    \begin{itemize}
    \item \textbf{Lack of personalization}: The existing solutions are not \textit{personalized} to the fund and hence require major manual work to generate added value for the funds. In other words, these solutions do not know the fund investment thesis and its unique strengths and requirements. Rather their outcome is generic. However, in reality, every fund is unique in its investment thesis, approach, and portfolio, and understanding such uniqueness is critical for driving value. It takes a large human effort to turn the outcome of these machines into real value for the fund.
    \item \textbf{Limited intelligence}: Lack of access to the fund data also results in limited intelligence and ability to learn. For a machine learning algorithm, e.g., a recommendation algorithm, to achieve competitive performance, it should have an active pipeline of data from the fund to grow more intelligent. For example, it should receive feedback from the fund, navigate which data it is missing to actively collect it, and even use the data available at the portfolio companies. Since none of these is available to third party software solutions, they fail to get more intelligent over time. 
    \item \textbf{Privacy concerns}: Providing the fund data to a third party comes at a large privacy risk. An example of such risks is uploading data gathered under NDA from a target asset to a third party database. In fact, this privacy concern is one reason why third party solutions can not become personalized. 
    \item \textbf{Natural limits of software}: Each software solution only covers part of this fragmented space and in an incomplete way. Software naturally is designed for certain use cases and cannot be extended to other use cases in the fund. This means that to address its diverse and emerging needs for automation, the fund will have to get access to a collection of software and try to put them all together. This will lead to cumbersome processes and inefficient outcomes.
\end{itemize} 

Due to the limits of third party software solutions for productivity tools, there has been a shift in the industry toward a \textit{data-driven platform solution}.

\section{What is the AI platform disruption in the investment business?}

Consequent to the shortcomings of the third party solutions, almost all major funds and some smaller funds have started building in-house proprietary automation platforms for their deal sourcing and deal insights: a game changer in the industry \cite{DataDrivenVC, PEHUB2022_1, BlackRock2023_1, PEI2023_2}. Some notable examples are the EQT Motherbrain machine, Blackstone data science team (BXDS), Blackrock Aladdin machine, Advent lab, and KKR data science team. These machines are primarily involved in recommending new deals and gathering deal insights and have already led many successful deals. There are also more than 80 VCs that have started building in-house automation machines which is expected to grow to 75\% of the VC funds by 2025 \cite{Gartner2021_1}. Similar trends exist in other types of investment funds.

These automation machines / AI platforms are different from traditional software solutions, in three important respects.
\begin{itemize}
    \item \textbf{Data integration platform}: These machines act as a platform to collect and connect all internal data and external data in one intelligent system. The machine has access to the fund's proprietary data, e.g. investment meetings or previous due diligence, and this way learns about the fund's unique approach to investment. What is important is that these machines gather more and more information about the fund over time, and this way become more intelligent about it. In this sense, every interaction of the investors with the AI machine makes them smarter. Furthermore, access to the selected data from portfolio companies, e.g. customer preference, sales, supply chain, etc., can help the machine to generate deep insights into their corresponding industry and trends.
    \item \textbf{Personalization platform}: As the machine learns about the fund, its outcomes are personalized to the fund. For example, the recommended deals for the fund are by considering the fund investment thesis, its unique strength and advantages in generating value from an asset, and the current state of the portfolio. Similarly, the deal insights identify the unique risks and upsides of a target asset by considering the fund's unique advantage.
    \item \textbf{Platform for diverse use cases}: The fund can use this platform to automate the different use cases all in a coherent way. This way, instead of multiple software, the fund has one machine that can grow over time to address different use cases. This is a particularly important feature with growing AI technology which can make automation of different fund functions feasible over time. Furthermore, each function can have multiple use cases which can be all integrated into this one internal machine.
\end{itemize}  

In understanding what the value driven by these platforms is, one should first note that the platforms are an \textit{augmentation to boost human performance}. Second, one should consider that the platform is for true data-driven use cases across \emph{all stages of the investment}.
\begin{itemize}
    \item First, the AI tool for investment is an augmentation to boost human performance. In other words, humans will work with the machine to form \emph{hybrid intelligence} in making the investment decisions. In fact, research has shown that a descent human augmented with a machine can outperform seasoned stars of the industry \cite{Augmentation_1}. By boosting human performance, from analysts to partners, the machine allows them to spend time on real value-added tasks i.e. building relationships and making investments.
    \item Second, the in-house proprietary AI machine provides a platform for the fund to derive and benefit from different use cases \emph{across the investment cycle}. Currently, these machines are primarily used for deal sourcing, including M\&A options, and deal insights which are hand-in-hand: any recommendation should be followed by clear insights and reasoning. As opposed to the third party software solutions, the use cases from these platforms are not limited to productivity tools and only at the beginning of the investment cycle e.g. database of companies to search or doing standard due diligence with a ChatGPT trained in-house. Some examples of true data-driven use cases for deal sourcing across the investment cycle are the following:
    \begin{itemize}
        \item The machine recommends proprietary deals or M\&A for the fund by leveraging mainstream investment data, e.g. Crunchbase, Pitchbook, Sourcescrub, etc, alternative data e.g. large customer review, sentiment, mobility, credit card, etc., and any proprietary data at the fund or its portcos. The recommendation can be by taking archetypes directly from investors, or more interestingly by freely exploring the space including adjacent sub-sectors to current investment space, and industries with transferable knowledge/expertise.
        \item The machine identifies the sub-sector of a selected asset and finds all similar or close-by companies as alternatives.
        \item The machine provides personalized scores, risks, and upsides about the recommended deals.
        \item The machine tracks the asset for recommending a buy time. It identifies lead sell indicators for making a move on buying a selected asset, e.g. new C-level, new fundraising, organizational changes, etc., generates frequent reports, and sends an alert on the correct time to buy using the lead sell indicators.
     \end{itemize} 
Similarly, some examples of data-driven use cases for deal insight across the investment cycle are the following:
\begin{itemize}
    \item The machine identifies personalized key features and red flags to focus on in assessing a target. These are by considering the fund capabilities, its portfolio, and investment thesis, and using mainstream, alternative, and proprietary data.
    \item The machine provides detailed insights on key features and key risks for assessing a target using a multi-faceted and robust approach e.g. using a structured approach i.e. document processing, as well as an unstructured approach by interpreting the features of the neural network used in the recommendation system.
    \item The machine structures the management presentation meetings for focused conversations and getting smart earlier in the process, by proposing a list of data requests and questions from the target.
    \item The machine intakes diverse private data provided by the target and updates the corresponding key features, risk factors, insights, and next steps with the management presentations
    \item The machine further intakes diverse private data provided by the target and updates the corresponding key features, risk factors, insights, and next steps with the management presentations.
    \item The machine provides a non-binary score/vote in the investment committee meeting for investment in an asset.
\end{itemize}
The above example use cases show that the AI machine provides support to the investors from the beginning step of the investment to the final decision-making.
\end{itemize}

It is important to note that the surge of these in-house AI platforms has been backed by AI technological progress for investment processes \cite{MIT2023_1}. There are more than 200 recent academic research papers, some directly supported by investment funds, that provide algorithms for the automation of the investment processes (for a literature survey see \cite{EQT2023_2}). These papers have proposed and successfully tested machine-learning algorithms, using neural networks, in predicting the success of a company, recommending investments for a specific fund, generating insights from alternative data, and reasoning about an investment. The rapid progress in this space has led to more complex algorithms that are able to model the entire industry ecosystem and recognize where in this space the fund investment thesis lies (e.g. see \cite{TrendAware, EQT2023_3, unicorn}). The success of these algorithms has provided sufficient proof of concept to ensure real value for the funds using them.

\section{How have large language models changed the game for AI platforms in the investment business?}
ChatGPT, PaLM,  Falcon, Cohere, and other similar open-source large language models have fast-forwarded the AI platform disruption in investment, and allow small and mid-size funds to also enter the game for recapturing their competitive advantage by developing in-house proprietary platforms. Despite the common perception, the real impact of these models is far beyond just using ChatGPT as a chatbot for asking standard due diligence questions regarding an asset. In fact, these models can be used in different parts of the AI platform to significantly improve its performance (for example \cite{Moonfire_1}). As such, the impact of these open-source large language models can be categorized into the following 6 areas.
\begin{itemize}
    \item \textbf{Data preparation/cleaning}: These large language models have significantly reduced the cost of data cleaning with their ability to structure unstructured data. Many funds do not have an organized and well-structured database to use for their platform, and the process of preparing data to be used for the machine learning algorithm is time-consuming and costly. Large language models can intake diverse data and put them into predefined structures.
    \item \textbf{Model training}: These models have reduced the cost and time required for training machine learning models for the investment use cases. Training and tuning a machine learning algorithm to achieve high performance for each use case is a complicated task. The large language models however have provided a base mega model that can be retrained for each specific use case in a faster and more cost-efficient way.
    \item \textbf{Proprietary data for personalization}: These models have reduced the amount of proprietary data required for driving value uniquely to the fund. While training a neural network model from scratch in a way to understand the fund's fingerprint requires a large amount of proprietary data, large language models can be retrained to provide even better insights with significantly smaller data.
    \item \textbf{Privacy preservation}: These models have provided privacy-preserving solutions. While third-party solutions are always at risk of data leakage, the open-source nature of the large language models allows forking them and using them in the fund servers. This way, the model and all the data associated with training it are kept at securely inside the fund.
    \item \textbf{Text embedding}: These models can significantly improve the quality of text embedding, a major function used in machine learning algorithms for investment processes. Text embedding allows turning text inputs about assets or funds into a numerical vector which can be then used by the machine. The quality of such embedding is a determining factor for the quality of the machine learning algorithm. Large language models are particularly good at processing text and generating such embedding.  
    \item \textbf{Interpreting neural networks}: These models can be used to interpret the features of a neural network, a function that is required for generating insights regarding the recommended deals. Neural networks, the main tool in AI engines, are black boxes that can recognize patterns in data. However, a major challenge in using neural networks is their lack of interpretability. In other words, the features selected by the neural network machine are not easy to interpret. This is while in the investment business, any recommendation should be supplemented with clear reasons and insights \cite{Interpretability_1}. Large language models have provided a new approach to interpreting the features of neural networks, which can be used for reasoning about the recommended deals and providing insights.
\end{itemize}
While the large language models are still in their early test phase for the investment industry,  they have already shown major improvements to the AI platforms and are promising more improvements in the future.

\section{Where should a fund start practically for an AI platform?}

For a practical approach to developing an in-house AI platform, a fund should answer six questions: 
\begin{itemize}
    \item What are the requirements from/features of its AI platform specific to the fund?
    \item What are the priority use cases for automation using the platform?
    \item What are the data fed into the AI platform?
    \item What are the models and algorithms used in the platform?
    \item Who is the development team i.e. build-vs-buy?
    \item What is the cost of development?
\end{itemize}
We elaborate on each of these six questions separately below.

\textbf{Features of the AI platform}: A fund should first define the set of features required from its platform to set the foundation of its development. While each fund's AI platform may have unique features, there are a few essential features that a successful AI platform should have. These features include adaptive learning, personalization, privacy-preserving, and forward compatibility. Adaptive learning ensures that the machine has an active pipeline of gathering valuable data, either internal or public, to enhance its intelligence. This includes receiving active feedback from the investors in the fund. Personalization ensures that in all use cases, the AI machine is delivering outcomes unique to the fund and matched to its investment thesis rather than generic outcomes. Privacy-preserving is to ensure the machine has clear boundaries on how it uses each piece of private data and ensures no leakage of the data either externally, or internally between different entities e.g. different assets sharing data under different NDAs. Finally, forward compatibility ensures the machine technology can adopt the latest developments in the AI industry, such as new advancements in the recommendation algorithms or new open-source foundation models.

\textbf{Priority use cases}: Second, the fund should select the priority use cases from its AI platform. There are two rules of thumb in choosing those use cases.
\begin{itemize}
    \item The fund should ensure it captures those low-risk high-return outcomes that are currently available with state-of-the-art technology. Deal sourcing and deal insight, both personalized to the fund, are priorities in this sense. As mentioned before, the current state of the technology and research in recommendation systems for investment have provided sufficient proof of concept for these two use cases. Furthermore, ChatGPT and other open-source large language models can improve the quality of these recommendation systems significantly. It is important that the fund correctly defines the corresponding use cases for deal sourcing and deal insight to garner the true value of AI from its platform. 
    \item The AI platform use cases should not be limited to early-stage asset discovery, and rather they should continue providing insights and supporting the investors through the process to the last stage of the investment decision. In fact, the AI machine is even more valuable in providing complementary insight to the target asset as the process gets closer to the final investment decision.
\end{itemize}

\textbf{Data fed into the AI platform}: The fund should ensure the machine has a continuous pipeline of data from diverse sources to achieve a competitive edge. Generally, the data sources include three types: a) public data including standard data, e.g. Crunchbase, Pitchbook, etc. and alternative data, e.g. reviews, social media, credit spent, etc., b) proprietary data including internal data and portfolio company data, and c) private data from its target assets provided under NDA. For each of these three sources, the fund should carefully investigate where the valuable data exists, and put in place a data-gathering pipeline to collect those data. It is important that the machine continuously receives more data from each source to enhance its intelligence and grow smarter about the investment ecosystem. Moreover, the machine should be constantly monitored for areas with a lack of sufficient data to be enhanced. This way the machine can achieve the highest competitive advantage. 

\textbf{Models and algorithms}: In choosing an engineering architecture and the corresponding models/algorithms for developing the AI platform, the fund should ensure the use of diverse algorithms and enhance them with the true power of open-source large language models e.g. ChatGPT. The investment use cases, such as deal sourcing and deal insight, are by nature multi-faceted tasks that should be done in a robust way; no single algorithm can always be the best. In this way, the fund should ensure the use of multiple diverse algorithms to generate the best outcome. Second, as mentioned above, the open-source foundation models are transformative to the performance of these algorithms if used correctly and at their full capacity.

\textbf{Development team i.e. build-vs-buy}: In selecting a team for building the AI platform, the fund should consider three capabilities: technical capability, know-how of investment processes, and ability to build along. The technical capability includes not only engineering skills but also advanced research. State-of-the-art AI architectures and algorithms for investment use cases are growing with many academic publications. A technical team should be able to connect to this body of research. The know-how of the investment processes is required for defining the correct and sharp use cases for adding value for investment. Finally, building such a platform requires working closely with the stakeholders in the investment fund. This includes defining the use cases, gathering feedback by interacting with the machine, developing the user interface, and training them to use the machine. Handling these interactions in a smooth and frictionless way to the day-to-day work of the fund is key to a successful build. A fund can either build a team in-house for this purpose or outsource the build. While building in-house can result in building the engineering capability internally, the outsourcing solution can be faster, at a lower cost, and with less friction for the fund. Another important benefit of outsourcing, especially for funds less than \$50B asset under management, is that the base platform infrastructure can operate as a public utility, with learnings and intelligence gains from all participant usage accruing to all end users.  This approach maximizes the velocity of learning and overall intelligence of the machine, while carefully enabling fund privacy and differentiation.  More on this will be described in the next paragraph.

\textbf{Cost of development}: While the cost of such an in-house bespoke AI platform is not comparable with SaaS solutions, considering larger funds are budgeting tens of millions of dollars for it, there are two factors that can reduce the costs for the new joiners. First, as mentioned above, the open source foundation models such as ChatGPT can reduce the time and cost of building a platform. Second, the engineering solution consists of three layers, two of which are fund-independent and can be shared across funds. These three layers are the infrastructure layer, the public data training, and the personalization layer. The infrastructure layer includes preparing the neural network algorithms for recommendation, graph modeling of the industry, data gathering pipeline, APIs for working with ChatGPT and other large language models, etc. The public data training layer includes training the models with investment knowledge including standard data and alternative data. The personalization layer is the main driver of value and includes installing the infrastructure at the fund, structuring its proprietary data, defining its platform features and use cases, training algorithms for its use cases, training employees, and embedding the solutions into the existing IT solutions. While the first two layers are fund-independent and can be shared across funds, the personalization layer is bespoke to the fund and adds the main value.

\section{How the risks can be evaluated in taking any AI response?}

The largest risk in taking an AI response is not taking any action. The industry shift has already started, most major funds have started developing in-house tools and many smaller funds are doing similar developments. The technology has passed the fermentation phase of its S-curve and is starting its take-off phase. In two years many funds will have such AI platforms in-house. Hence, while it is not a competitive disadvantage for a fund to not have such an AI platform now, it will be one in two years \cite{MIT2023_1}.

On the other hand, the early movers will stay ahead of the curve in this space. Unlike other technology sections where early movers face an exploration risk, in this AI space, early movers have an advantage because of the personalized nature of the solutions. It takes time for a fund to prepare the infrastructure for its AI response, clean its historical data, learn its unique way of using AI, and have the cultural transformation in using the technology. While the large language models have made it far more accessible for new funds to enter this game, in two years it may be too late to enter the competition to recapture the competitive advantage.

One major category of risk in using AI tools is privacy and compliance. As mentioned above, an AI platform should have the privacy-preserving feature. There are three standard practices for this purpose: First, the fund should develop its in-house machine whenever training proprietary and private data. The in-house machine is kept on the fund servers securely to stop external leakage of the data. Second, the fund should train separate models with clear boundaries when working with the data from different assets under different NDAs. The models should consequently be deleted after the NDA timeline terminates. This practice forbids using data from one asset to generate insights on another asset. Finally, for highly sensitive data, e.g. HR data, the fund may consider not training the model directly on those and rather just use the machine to search through those documents for required outcomes. This way it is clear how the sensitive data of each asset is used within its own insights. While these three are the basic practices, there are more privacy and compliance practices that a fund should consider.

\section{Conclusion: What is the future of AI in investment?}

The AI platform disruption has already started in the investment sector, now at the end of the fermentation phase of the S-curve, and beginning of the takeoff phase. Most major funds and many smaller funds have already started developing such in-house proprietary AI platforms. The recent developments in the large language models, e.g. ChatGPT, have made it easier for other funds to enter the game. While it is not a competitive disadvantage for a fund to not have an AI platform now, it will be in two years. 

For an investment fund, deal sourcing and deal insights are the first functions to start automating. However, there are many other functions across the investment cycle that can be automated with such a proprietary platform. For example, the graph model of the entire investment ecosystem can be used to recommend investors to the funds, individuals to the portcos, etc. The same 360-degree view of the industry can be used for validating and updating the fund investment thesis, running a portfolio simulation, and recommending paths for portfolio correction. Moreover, the prediction engine can be used to predict the price of a deal. The funds can use their AI platform to recommend transformation for their portfolio companies, support them with automation use cases, and monitor their progress. For another example, the fund can use the automation machine to predict the correct exit time of an asset, recommend a list of potential buyers, and automate the vendor due diligence process. These are just some selected use cases for the future. Those funds that put the correct foundation for such an AI platform internally will garner large advantages in the short and long run.

\vspace{20pt}

\textbf{Authors' biography:}

\begin{figure}[H]
\includegraphics[width=2cm]{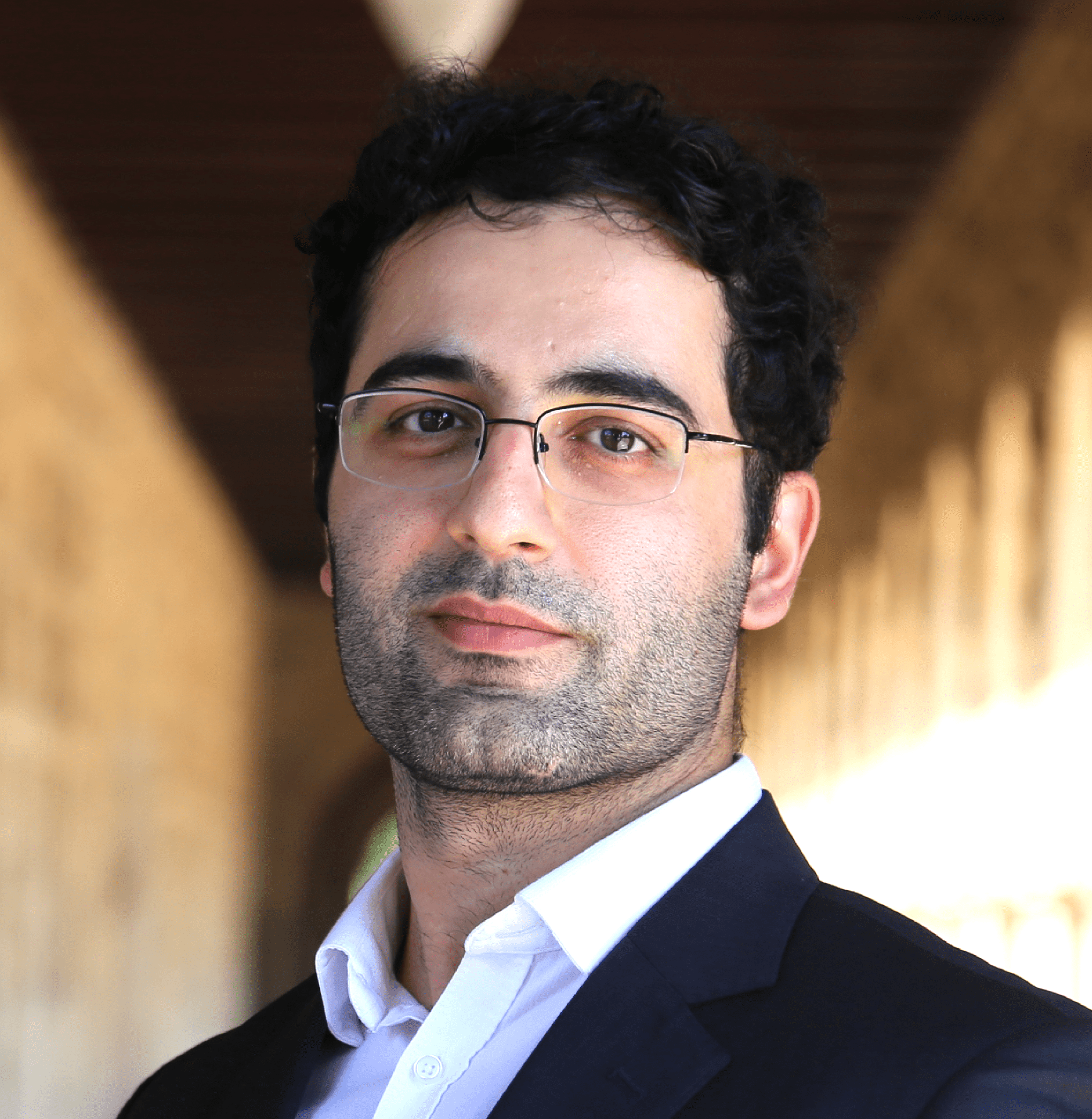}
\end{figure} 
\vspace{-15pt}
\textbf{Dr. Mohammad Rasouli} is a thought leader in AI and technology for investment. He is an ex-McKinsey consultant from the Bay Area and the New York offices managing AI activation projects for private equity clients. He has worked with top-20 private equities as well as a number of middle and small-size PEs to use AI technology for automating their fund processes as well as to use AI for their portfolio companies. Before joining McKinsey, Mohammad was an AI researcher at Stanford University publishing papers in top AI conferences with 450+ citations. He has particularly studied recommendation systems, working closely with companies. He has also co-taught “Empirics of Online Marketplaces” at Stanford Graduate School of Business. He finished his Ph.D. in Electrical and Computer Engineering, and his Masters in Economics from the University of Michigan working with a thesis committee of Stanford, Harvard, and MIT professors, including Nobel prize winners. Mohammad has also been a Microsoft engineer and has experience in startups including CTO and CEO roles.

\vspace{20pt}

\begin{figure}[H]
\includegraphics[width=2cm]{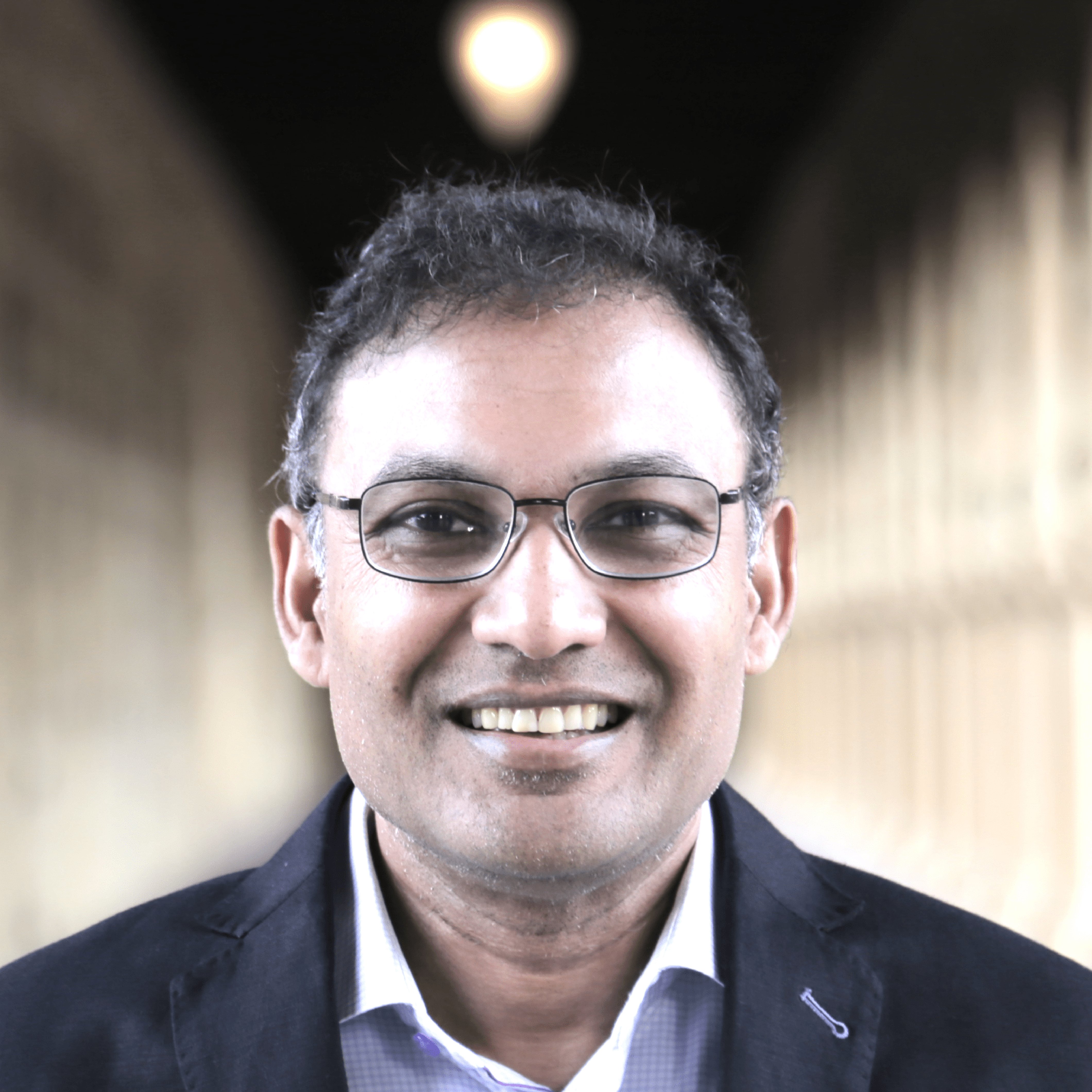}
\end{figure} 
\vspace{-15pt}
\textbf{Ravi Chiruvolu} is a technology visionary and provides a unique perspective of being a MIT trained AI engineer and a top quartile private company investor for 25+ years.  While at MIT and NASA, he helped create expert systems and a visual display systems for complex tasks (remotely controlling robots despite limited inputs), so maximizing the opportunity to apply AI platforms to investing to create sustainable competitive advantage is something he intuitively understands. Ravi has a BS and MS from MIT, worked at multiple NASA facilities, was a Business Analyst at McKinsey, a corporate strategist at Ameritech, and a General Partner and Managing Partner of two Investment Funds (Alta Partners and Charter Ventures). Ravi has an MBA from Harvard Business School and is a Retired First Lieutenant in the US Army. 

\vspace{20pt}

\begin{figure}[H]
\includegraphics[width=2cm]{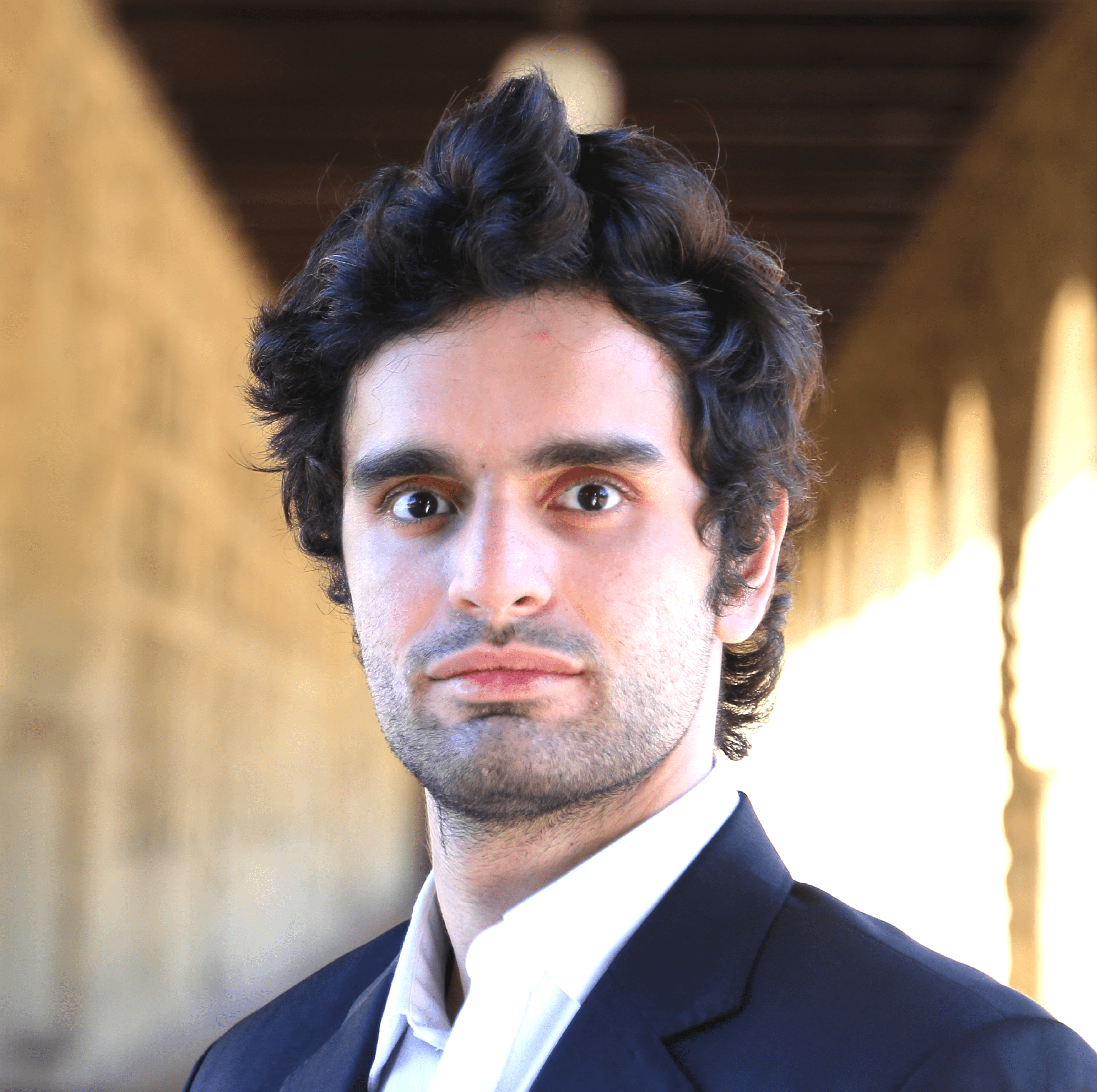}
\end{figure} 
\vspace{-15pt}
\textbf{Ali Risheh} is an AI researcher and engineer. His research has been in the field of large language models and neural networks, particularly graph convolutional neural networks. He has collaborated with researchers from Stanford University, Virginia Tech, and the University of Toronto. He has studied MSc in Computer Science at California State Los Angeles University. Ali has been a prompt engineer, also CTO and co-founder of machine learning startups.

\input{}



\bibliographystyle{unsrt}
\bibliography{main}


\end{document}